# Theoretical Study of Isotope Production in The Peripheral Heavy-ion Collision $^{136}Xe + Pb$ at 1 GeV/nucleon

H. Imal[*], R. Ogul

*Department of Physics, Faculty of Science, University of Selçuk, 42079, Konya, Turkey*

**Abstract**

We have studied the fragment yields emitted from the fragmentation of excited projectile nuclei in peripheral collisions of $^{136}Xe+Pb$ at 1 GeV/nucleon, and measured with the high-resolution magnetic spectrometer, the Fragment Separator (FRS) of GSI. The mass, charge and isotope distributions of nuclear fragments formed in the reactions were calculated within a statistical ensemble approach and compared to the experimental data. The ensemble of excited projectilelike source nuclei were created in the framework of a previous analysis of similar reactions performed at 600 MeV/nucleon (ALADIN-experiments, GSI). In addition, the ensemble of the low-excited compound nuclei is involved in the analysis. The overall agreement between theory and experiment was very satisfactory in reproducing the experimental data of isotope yields measured in the heavy-ion collisions. It is confirmed that a broad range of elements can be reproduced within a universal statistical disintegration of the excited projectile residues.

## 1. Introduction

Radioactive isotopes are produced mainly in research reactors, accelerators, and separation facilities. They are widely used in various fields including medicine, research, industry, and radiation processing of foods. This paper mainly aims to search for the isotope yields of nuclear fragments produced in high energy heavy-ion reactions. When a large amount of energy is deposited in a nucleus, the excited nuclear matter disintegrates into several fragments. It is called multifragmentation if there are at least 3 intermediate mass fragments (IMFs) with $Z \geq 3$ in one fragmentation event. It has been observed in nearly all types of high-energy nuclear reactions induced by hadrons, photons, and heavy ions (see, for example, Refs. [1-5]). At low excitation energies ($Ex \leq 1$ MeV/nucleon), the produced nuclear system can be treated as a compound nucleus [6], which decays via evaporation of light particles or fission. However, at high excitation energy, possibly accompanied by compression during the initial dynamical stage of the reaction [5, 7-9] the system will expand to subsaturation densities, thereby becoming unstable, and will break up into many fragments.

At sufficiently high projectile energies, from a few tens of MeV per nucleon up to the order of GeV, all stable and radioactive isotopes starting from hydrogen including its well-known radioisotope tritium, up to the heaviest nuclide can be produced in nuclear multifragmentation reactions [10, 11]. As we move away from the stability valley, these isotopes the so-called exotic nuclei become even more difficult to produce, separate and identify, due to their very low production cross-sections and very short lifetimes, and the large number of other unwanted isotopes produced in the targets. However, their separations and identifications have become possible with the development of advanced technologies such as ISOL (Isotope Separation On-Line) and In-Flight Separation techniques. These accelerator-based processes are more difficult and complex than the activation processes for producing isotopes in nuclear reactors [10-14]. The nuclear multifragmentation studies are even more important in modelling astrophysical events, and for practical applications in medicine, space research and technology.

[*] Corresponding author. Tel.: +0-000-000-0000 ; fax: +0-000-000-0000 ; e-mail: himal@selcuk.edu.tr

Existing results in the literature and our recent analyses show that the statistical models are very successful for interpretation of nuclear reactions in a wide range of energy. They are used for description of nuclear decay when an equilibrated source can be identified in the reaction. In the most general consideration, the decay process may be subdivided into several stages: (1) a dynamical stage leading to formation of equilibrated nuclear system, (2) disassembly of the system into individual primary fragments, (3) de-excitation of hot primary fragments. More detailed information on how statistical approaches is applied to the analysis of nuclear multifragmentation reactions can be found in the recent studies and references therein [15-20].

In the present paper, we have theoretically analysed the experimental data of projectile fragments produced in the fragmentation reactions $^{136}$Xe + Pb, at an incident beam energy of 1 GeV/nucleon. Production cross-sections of projectilelike isotopes were measured with the high-resolution magnetic spectrometer, the Fragment Separator (FRS) of GSI, Darmstadt [21]. The heavy-ion beams were delivered from the universal linear accelerator (UNILAC) to the SIS18 heavy-ion synchrotron, where they were extracted and guided through the target area to the FRS. The FRS was used for the separation and analysis of the isotopes emitted from the excited projectiles. For the interpretation of experimental data, we have carried out the calculations in the framework of the statistical multifragmentation model (SMM) for nuclear multifragmentation process.

## 2. Statistical simulation of the formation of radioactive isotopes

There are several models in the literature developed to analyse high energy nuclear collisions at various energies. At low excitation energies up to $E_x$ = 1 MeV/nucleon, disintegration of heavy nuclei is very well described by the compound nucleus model introduced by Niels Bohr in 1936 [6]. At higher energies, besides the dynamical and kinetic models, the statistical models are found to be very suitable to reproduce the experimental data, especially in describing the fragment production [15, 22]. In this study, we consider the statistical ensemble version of SMM. The same parameters that were used for a previous analysis of ALADIN experimental data obtained for similar reactions at 600 MeV/nucleon [15], were considered. The general properties of the considered ensembles of residual nuclei are given in Ref. [23].

*2.1. Description of the statistical model*

The SMM assumes statistical equilibrium of the excited nuclear system with mass number $A_0$ charge $Z_0$, and excitation energy $E_x$ (above the ground state) within a low-density freeze-out volume. In the SMM, all breakup channels (partitions j) composed of nucleons and excited fragments are considered and the conservation of baryon number, electric charge number, and energy are considered. Besides the breakup channels, also the compound-nucleus channels are included and the competition between all channels is permitted. Thus, the SMM covers the conventional evaporation and fission processes occurring at low excitation energy as well as the transition region between the low- and high-energy de-excitation regimes. In the thermodynamic limit, as demonstrated in Ref. [24], the SMM is consistent with the nuclear liquid-gas phase transition when the liquid phase is represented by an infinite nuclear cluster.

In the model, light nuclei with mass number $A \leq 4$ and charge $Z \leq 2$ are treated as elementary stable particles with masses and spins taken from the nuclear tables ("nuclear gas"). Only translational degrees of freedom of these particles contribute to the entropy of the system. Fragments with A>4 are treated as heated nuclear liquid drops. Their individual free energies $F_{A,Z}$ are parametrized as a sum of the bulk, surface, Coulomb and symmetry energy contributions, respectively, as follows:

$$F_{AZ} = F_{A,Z}^B + F_{A,Z}^S + E_{A,Z}^C + E_{A,Z}^{sym} \tag{1}$$

The standard expressions for these terms are

$$F_{A,Z}^B = (-W_0 - T^2/\varepsilon_0)A \tag{2}$$

where, $T$ is the temperature, $\varepsilon_0$ is related to the level density, and $W_0 = 16 MeV$ is the binding energy of infinite nuclear matter,

$$F_{A,Z}^S = B_0 A^{2/3}((T_C^2 - T^2)/(T_C^2 + T^2))^{5/4} \tag{3}$$

where $B_0 = 18 MeV$ is the surface energy coefficient, and $Tc$ = 18 MeV is the critical temperature of infinite nuclear matter;

$$E_{A,Z}^C = c\, Z^2/A^{1/3} \tag{4}$$

where, we have

$$c = (3/5)(e^2/r_0)(1 - (\rho/\rho_0)^{1/3}) \tag{5}$$

which is the Coulomb parameter (obtained in the Wigner-Seitz approximation) with the charge unit $e$ and $r_0 = 1.17$ fm;

$$E_{A,Z}^{sym} = \gamma(A - 2Z)^2/A \tag{6}$$

where $\gamma$ = 25 MeV is the symmetry energy parameter. These parameters are those of the Bethe-Weizsäcker formula and correspond to the assumption of isolated fragments with normal density in the freeze-out configuration, an assumption found to be quite successful in many applications. However, these parameters, especially the symmetry coefficient $\gamma$, can be different in hot nuclei at multifragmentation conditions, and they should be determined from corresponding experimental data as shown throughout the recent studies by various groups [15, 16, 19, 25]. According to the microcanonical treatment [1], total number of nucleons, total charge and energy are fixed, and the statistical weight of a partition j is calculated by

$$W_j = \frac{1}{\xi} exp\,(S_j(E^*, A, Z)) \tag{7}$$

where, $\xi$ is the normalization constant (the partition sum), given by

$$\xi = \sum_j exp\,(S_j(E^*, A, Z)) \tag{8}$$

and $S_j$ is the entropy of the channel j, depending on the fragments in this partition as well as on the parameters of the system. In the canonical ensemble the temperature, the total nucleon number and charge are supposed to be fixed for all partitions. The probabilities of different break-up channels are determined in terms of Gibbs free energy and temperature, instead of the entropy of the channel in the microcanonical treatment [1, 2]. At low energy excitations, the decay channels of compound nucleus are also included and the competition between all channels is permitted. Thus, the conventional evaporation and fission processes in the transition region between the low and high energy de-excitation regimes are also included in this model. Further details of the model can be found in Ref. [2].

Here we particularly stress two main achievements of statistical models in theory of nuclear reactions: first, a clear understanding has been reached that sequential decay via compound nucleus must give a way to nearly simultaneous break-up of nuclei at high excitation energies; and second, the character of this change can be interpreted as a liquid-gas type phase transition in finite nuclear systems. The results obtained in the nuclear multifragmentation studies can be applied in several other fields. First of all, the mathematical methods of the statistical multifragmentation can be used for developing thermodynamics of finite systems [26]. These studies were stimulated by recent observation of extremely large fluctuations of energy of produced fragments, which can be interpreted as the negative heat capacity [27, 28]. A very important advantage of the statistical approach to the cluster production is that the statistical equilibration is generally achieved in the astrophysical conditions [29]. We can demonstrate the links for the neutron-rich isotope production in both cases in Refs. [30, 31]

*2.2. De-excitation of hot fragments*

For excitation energies $E_x \leq 1$ MeV/nucleon, the secondary excitation of the hot particles can be simulated according to the SMM code within the standard Weisskopf evaporation and fission scheme [32]. The decay of light fragments $A \leq 16$ can be described by the Fermi break-up model. In the microcanonical approximation we consider all possible breakup channels satisfying the conservation of energy, momentum, and particle numbers A and Z. We assume that the probability of each event channel is proportional to the occupied states in the phase space. The weight of the channels containing *n* particles with masses $m_i$ ($i = 1, \cdots, n$) is given by

$$W_j^{mic} = \frac{S}{G} \left(\frac{V_f}{(2\pi\hbar)^3}\right)^{n-1} \left(\frac{\Pi_{i=1}^n m_i}{m_0}\right)^{3/2} \cdot \frac{(2\pi)^{\frac{3}{2}(n-1)}}{\Gamma\left(\frac{3}{2}(n-1)\right)} \cdot (E_{kin} - U_j^C)^{\frac{3}{2}n - \frac{5}{2}} \qquad (9)$$

where $m_0 = \sum_{i=1}^n m_i$ is the total mass, $S = \Pi_{i=1}^n (2S_i + 1)$ is the spin degeneracy factor ($S_i$ is the i-th particle spin), $G = \Pi_{j=1}^k n_j!$ is the particle identity factor, $n_j$ is the number of particles of kind j, $E_{kin}$ is the kinetic energy of nuclei and $U_j^C$ is the Coulomb interaction energy between nuclei, which are related to the energy balance as described in Ref. [33]. The total excitation energy $E_x$ can be expressed using the conservation of total energy.

Sequential decay modes of primary hot fragments with mass number $A > 16$ were studied nearly 60 years ago as excited modes of compound nuclei [32]. This mechanism has been investigated extensively, and it was shown that compound nucleus models successfully reproduce the experimental data [2]. The emission width of a particle j emitted from the compound nucleus (A, Z) is given by

$$\Gamma_j = \sum_{i=1}^n \int_0^{E_{AZ}^* - B_j - \epsilon_j^{(i)}} \frac{\mu_j g_j^{(i)}}{\pi^2 \hbar^3} \sigma_j(E) \frac{\rho_{A'Z'}(E_{AZ}^* - B_j - E)}{\rho_{AZ}(E_{AZ}^*)} E dE \qquad (10)$$

Here the sum is taken over the ground and all particle-stable excited states $\epsilon_j^{(i)}$ (i=0,1,...,n) of the fragment j, $g_j^{(i)} = (2S_j^{(i)} + 1)$ is the spin degeneracy factor of the i-th excited state, $\mu_j$ and $B_j$ are corresponding reduced mass and separation energy, $E_{AZ}^*$ is the excitation energy of the initial (mother) nucleus, and $E$ is the kinetic energy of an emitted particle in the centre-of-mass frame. In Eq. (10), $\rho_{AZ}$ and $\rho_{A'Z'}$ are the level densities of the initial ($A$, $Z$) and final (daughter) $(A', Z')$ compound nuclei in the evaporation chain. The cross-section $\sigma_j(E)$ of the inverse reactions $(A', Z') + j = (A, Z)$ was calculated using the optical model with nucleus-nucleus potential [1]. This evaporation process was simulated by the Monte Carlo method and the conservation of energy and momentum was strictly controlled in each emission step. After the analysis of experimental data, we come to conclusion that at sufficient large excitation energies (more than 1 MeV per nucleon) it is reasonable to include the decreasing the symmetry energy coefficient in mass formulae, that leads to adequate description of isotope distributions [16,19, 34].

An important process of de-excitation of heavy nuclei (approximately, $A \geq 100$) is the fission of nuclei. This process competes with particle emission, and it can also be simulated with the Monte-Carlo method at each step of the evaporation-fission cascade. Following the Bohr-Wheeler statistical approach we assume that the partial width for the normal compound nucleus fission is proportional to the level density at the saddle point $\rho_{sp}(E)$, as follows:

$$\Gamma_f = \frac{1}{2\pi\rho_{AZ}(E_{AZ}^*)} \int_0^{E_{AZ}^* - B_f} \rho_{sp}(E_{AZ}^* - B_f - E) dE \qquad (11)$$

where $B_f$ is the height of the fission barrier which is determined by the Myers-Swiatecki prescription. For approximation of $\rho_{sp}$ we have used the results of the extensive analysis of nuclear fissility and branching ratios $\Gamma_n$ / $\Gamma_f$ (see Ref. [2]).

## 3. Calculations and comparisons with experimental data

In this section, we compare the results of our calculations on reproducing the fragment mass and charge yields in the reactions $^{136}$Xe+Pb at 1 GeV/nucleon, and the experimental data obtained for the same reaction system in Ref. [21]. In the calculations, the statistical ensemble version of SMM was performed to describe the properties of isospin asymmetric reaction system (projectile and target nuclei have different neutron to proton ratio N/Z). The primary hot fragments emitted from excited quasi-projectile sources were assumed to be formed during the preequilibrium process long before the statistical equilibrium stage. The de-excitation process leading to the formation of final cold fragments starts after the disintegration of the system into primary hot fragments. The present analysis is based on Ref. [23], where the distribution of excited projectile sources were characterized by a correlation of decreasing source mass number A corresponding to increasing excitation energy per nucleon $E_x/A$ and by a saturation of excitation energy per nucleon at around $E_x/A \approx 8$ MeV/nucleon.

*3.1. Mass and charge distributions*

For a quantitative comparison of theory and experiment, normalizations have been carried out with respect to the measured cross-sections in the interval $30 \leq Z \leq 35$, as shown in Fig.1. We obtained the factor 0.036 mb per theoretical event for fragmentation of $^{136}Xe$ projectiles. In our previous study [15], and [23], the parameter dependence of the yields for ALADIN experiments was investigated in detail. According to these analyses, the distributions of excited sources are characterized by a correlation of decreasing mass number of the sources $A_s$ with increasing excitation energy per nucleon $E_x/A$ and by a saturation of the excitation energy at values between $E_x/A \approx 6$ to 8 MeV. It was seen that the excitation energy of the ensemble sources increase with decreasing values of the parameter $a_2$. To optimize the present calculations with the same parametrization, we have obtained a good fit to the experimental yields given in Ref. [21], as well.

In Fig.1, the upper panel shows the theoretical results obtained from ensemble calculations performed for 100 000 reaction events for mass distribution of final cold fragments, and the experimental data given in Ref. [21]. The lower panel shows the charge distribution as function of charge number Z, for the region of $A \geq 20$, $Z \geq 10$. In the experiments, the production cross-sections were restricted to the final cold residues in the charge number range $Z \geq 10$ since the correction for the limited angular transmissions through the FRS was performed for $Z \geq 10$ (see, Ref. [21]). This is because, the determination of the transmission corrections for lighter isotopes is rather difficult due to the overlap of contributions from different reaction mechanisms producing these light isotopes.

As a result, the U-shape of nuclear mass and charge distributions shown in Fig.1 evolve in the region between $Z \geq 10$ and the heaviest residues close to the projectile. Theoretical mass and charge distributions follow a similar trend to experimental ones measured in reactions with considerably high excitation energies. Theoretical results of these distributions (for Z<10) are not shown in Fig.1, since the experimental data are not available in this region of lighter particles. In short, we have shown that the mass and charge numbers of the particles formed in the reactions decrease as the excitation energy accumulated in projectile sources increases.

[Table 1 about here.]

In Fig.1, the most prominent difference between the theoretical predictions produced with ALADIN ensemble parameters (empty circles, in Fig.1) and the experimental data is observed for the yields of near-projectile elements in the vicinity of Z=54. This is because, these compound nuclei are formed by the weakly excited projectile sources in the region of very peripheral collisions, and they were not present in the ALADIN ensemble according to the conditions of the experiment. Since the residue ensemble for ALADIN is designed to calculate the data on multifragmentation, and to show the production of several nuclei (with $Z \leq 25$) is possible, while such a nucleus can be measured only through the FRS experiments. Hence, for the completeness of the paper, we have demonstrated that the yield of these big residues can also be reproduced within SMM by including the statistical weight of low excitations (compound nuclei with $E_x < 1$ MeV/nucleon) in the whole weighting. In order to include the contribution of compound nuclei, we have choosen, in addition, the interval of $E_x$ from 0 to 4 MeV per nucleon with the corresponding masses and charges of the nuclei. The statistical weight of the sources were taken over all ensemble according to their excitation energies that are assumed to be inversely proportional to the corresponding impact parameters.

[Figure 1 about here.]

In Table 1, we present the average parameters of this compound nuclei ensemble prepared with respect to the excitation energy of projectile residues, and the results were shown by blue dashed lines, in Fig.1. It is obtained after the estimate of the compound nuclei compositions and their probabilities made with Dubna Cascade Model (DCM) calculations [35]. The decreasing of N/Z ratio is related to the larger lost of neutrons in such collisions because of these neutrons are considerably on the nucleus periphery in such neutron rich nuclei. For charge yield in lower panel, we weighted the sources in the interval Ex=1-4 MeV/nucleon with binnings of 1 MeV, for which we were able to reproduce residual nuclei up to Z=54, and charge yields were consistent with the data. However, for the mass yield in upper panel, we managed to reproduce the heavy nuclei only up to A ≈ 124, within the interval Ex=1-4. Therefore, we had to consider the values of Ex<1, too, in the interval Ex=0.05-1 MeV/nucleon (i.e., over the 20 sequential values of Ex) with binnings of 0.05 MeV, in order to be able to reproduce heavy residues in the vicinity of the projectile (A=124-136 interval). Thus, we were able to reproduce the yields in the low excitation region, too, in the framework of the standard Weisskopf evaporation scheme in the SMM code, defined by Eq. (9), which are compatible with experimental data. In the lower excitations (approximately for Ex ≤ 0.1 MeV/nucleon) we noticed only the small oscillations of the compound nuclei without releasing any nucleons.

The predicted mass and charge yield results at low excitation region can be regarded as satisfactory for our current purposes, since they follow the same trend with the experimental data. For more rigorous calculations, the grazing angle should be simulated within a dynamical approach for the description of weakly excited systems. Because, the weakly excited nuclei (with Ex ≤ 0.1) are mostly in a vibration channel like ripples on the still water surface, without evaporation. In this case, the particles detected are the projectile nucleus itself. However, this is beyond the scope of the present paper.

Another feature striking in Fig.1 is that one may notice the distributions at Z=55 and 56 (greater than projectile charge number Z=54), which were detected in the experiments [21]. This is the so-called double charge-pickup residues assumed to form as a result of quasi-elastic collisions between a proton and a neutron of the target and the projectile source, respectively (i.e., a proton replaces the neutron inside the projectile-like fragments, that is the projectile would have lost a proton), or through the subsequent decay of the excited delta-resonance state of a projectile nucleon or a target nucleon.

Consequently, a qualitative agreement between theoretical and experimental results with considerable differences has been noticed. The observed differences in the distribution plots are mostly originated from the fact that the isotopes are not fully covered in the experiments. An extensive analysis of these differences was given in Refs. [15,19] for the relativistic heavy-ion collisions measured at FRS. In the Fermi energy region, where the projectile velocity is comparable with the Fermi velocity, nucleon exchange between the reaction partners during the collision become important. In the dynamical stage of the collision, N/Z values of quasi-projectile and quasi-target sources move towards the N/Z of composite system as a result of isospin diffusion, that produces significant changes in fragmentation picture (see, e.g., Refs. [36,37]) at Fermi regime.

According to the DCM, the loss of neutrons is greater than the loss of protons as a result of the pre-equilibrium emission. Therefore, in the selection of excited quasi-projectile sources in Table 1, we have considered that the N/Z ratios of the sources are not greater than that of the the projectile one and, moreover, the isospin equilibration process is also unlikely at relativistic collisions. This is supported by INC or BUU calculations which predict a slight decrease of the neutron richness with increasing excitation energy, at relativistic energies (see, e.g., Ref. [15]).

*3.2. Isotope distributions*

Isotope distribution observables may provide information in isospin asymmetry dynamics of the reaction systems. In Fig.2, we show the experimental data (red solid circles) of the isotope distributions of some selected ones given in Ref. [21] and compared to the presently predicted results in the framework of the statistical ensemble approach. In this figure, the results reproduced with the ALADIN parametrization were shown with solid lines, and in addition the results for weakly excited compound nuclei were shown with blue dotted lines in the last column for heavier fragments with Z=38,39,40. The full isotope distribution inside the angular acceptance of the FRS was applied for the experimental measurements [21]. It is seen from this figure that there is a slight discrepancy between predictions and experimental data. This is because, our recent analysis of ALADIN data shows that the isotopic production cross-section values (isotope distributions) are more sensitive to the variation of symmetry energy parameter at low density freeze-out region. For this reason, the best agreement with experimental data was achieved by taking the reduced symmetry term coefficient into account instead of standard value of 25 MeV. In the present calculations, we have taken the modified values of symmetry energy parameter obtained for a similar reaction system in Ref. [19]. Moreover, the calculations with the normal symmetry energy parameter (25 MeV) for these species will lead to more neutron poor isotopes that produce considerable changes in isotope curves. As clear from the references cited above and other similar ones in the literature, the reduced symmetry energy is actual only for multifragmentation events. For detailed information on such optimization calculations through the modifications of these model parameters, we will refer the reader to Refs. [15, 16, 36].

One may also see the production cross-sections of some selected radionuclides in Fig.2, whose names are highlighted in red, such as $^{18}$F (for use as a tracer in PET scans), $^{40}$K (for use of dating purposes, and as the largest source of natural radioactivity), $^{41}$Ca (for use of dating of carbonate rocks), and $^{90}$Sr (as one of the most dangerous fission product of nuclear fallout). As can be seen from the figure, all these isotopes have lower cross-section values, in other words, lower lifetimes than those of the most stable isotopes located around the maximum point of the isotope distribution curves. For example, one may see from the figure that positron emitter $^{18}$F, which is used for medical PET imaging, has a cross-section of 1.94 mb, while that of the most stable one $^{19}$F located at maximum, is 7.04 mb. Moreover, isotopes with a cross-section of less than about 1 mb probably have lifetimes in the order of milliseconds and therefore cannot be directly observed, but their decay channels can be measured [20].

[Figure 2 about here.]

## 4. Conclusions

In conclusion, we have analysed FRS data on nuclei and isotope production in relativistic heavy-ion collisions to demonstrate that it is possible to reproduce the yields with the previously adopted SMM approach. Including the compound nuclei contribution at low excitation energies, we have extended the SMM residual nuclei ensemble to the low excitation energy regime.

Apart from the producing isotopes in research reactors for medical and industrial use, it is also made for research purposes at low and high energy accelerators by various facilities. The chemical isotopes are mainly produced in stellar evolution including supernova explosions and formation of stars. There may be several elements occur in universe even though they cannot be found on earth, and some of them may be detectable in cosmic rays. Various research groups at the several accelerator facilities are trying to produce new elements/isotopes. In some laboratories including GSI, JINR, and RIKEN new elements in between Z=107 (Bohrium-Bh) and Z=118 (Oganesson-Og) were discovered in the period of the years 1981 and 2006 [38]. They are mostly unstable and decay after a short time in milliseconds, therefore they are not directly observed, but their decay chains can be measured. However, theoretical calculations have predicted the island of stability, whereby the isotopes of superheavy elements might have considerably longer lifetime [39].

So far, thousands of radioactive isotopes are predicted to lie within the particle stability limits in nuclear chart, and around fifty per cent of them are identified. Although, the Radioactive Ion Beams (RIB) facilities contribute substantially to work in this area, more work would be worthwhile to produce exotic isotopes contained in the chart. As a result of many years of research in large ion-beam accelerator systems it was also succeeded to develop a new form of radiotherapy the so called heavy-ion and/or hadron therapy. The advantage of this new treatment modality is that the ion beam selectively damages tumour tissues while sparing the surrounding healthy tissues.

Consequently, we show that neutron-rich radioactive isotopes can be produced in nuclear reactions in a wide range of energy regime. The main examples of such reactions are fission, fusion, nuclear fragmentation, and nucleon transfer reactions [10, 40]. At high energy heavy-ion collisions starting from the Fermi energy regime (20-50 MeV/nucleon) up to relativistic energies (1 GeV/nucleon), nuclear multifragmentation reactions become superior to produce neutron-rich and proton-rich isotopes [41, 42]. Further experiments are needed to extract information for the properties of neutron-rich exotic nuclei towards neutron dripline in nuclear chart. This kind of studies will also provide us with very useful tools for investigating the properties of stellar matter at extreme conditions, because of the similarities of nuclear and stellar matter.

**Acknowledgements**

The authors gratefully acknowledge enlightening discussions with A.S. Botvina and W. Trautmann on the interpretation of FRS and ALADIN experiments.

**List of Figures**

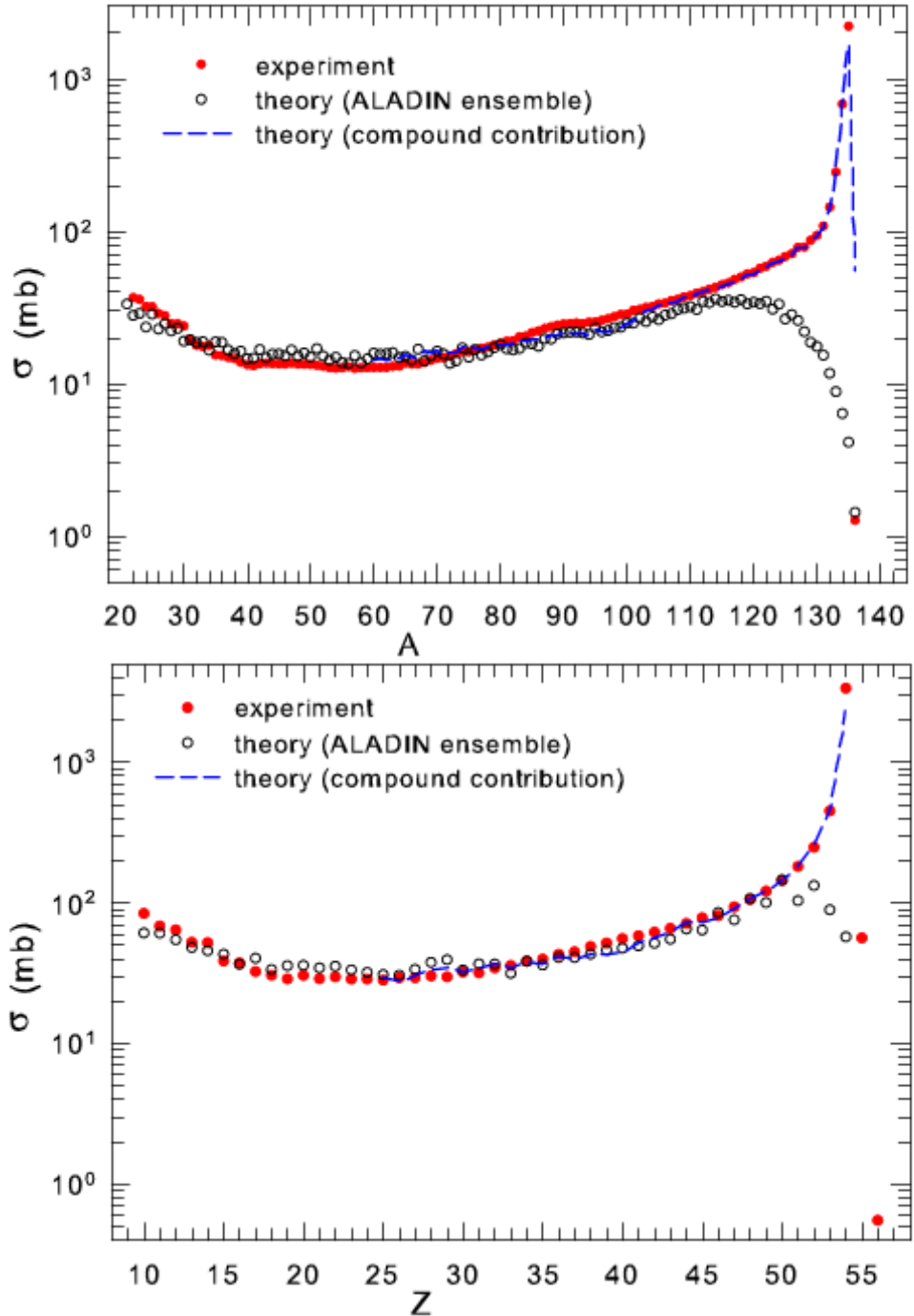

**Figure 1.** Mass and charge distributions of isotopes produced in the projectile fragmentation of the reaction system $^{136}Xe + Pb$. Upper panel shows the production cross-sections of the produced isotopes as a function of the isotope mass number A, and lower panel as a function of the isotope charge number Z. In both panels, the red solid circles show the experimental data, and the empty circles the theoretically predicted results within ALADIN parameters (see, Ref. [15]). The blue dashed lines show the results produced by statistical weighting taken for Ex = 1, 2, 3, 4 MeV/nucleon, including low excitations for Ex < 1. In the charge yield calculations (lower panel), statistical weights are taken in Ex =1 - 4 MeV/nucleon, while for the mass yield calculations (upper panel) the values of Ex < 1 are also taken into account to be able to reproduce the heaviest residues in the A=125-136 interval.

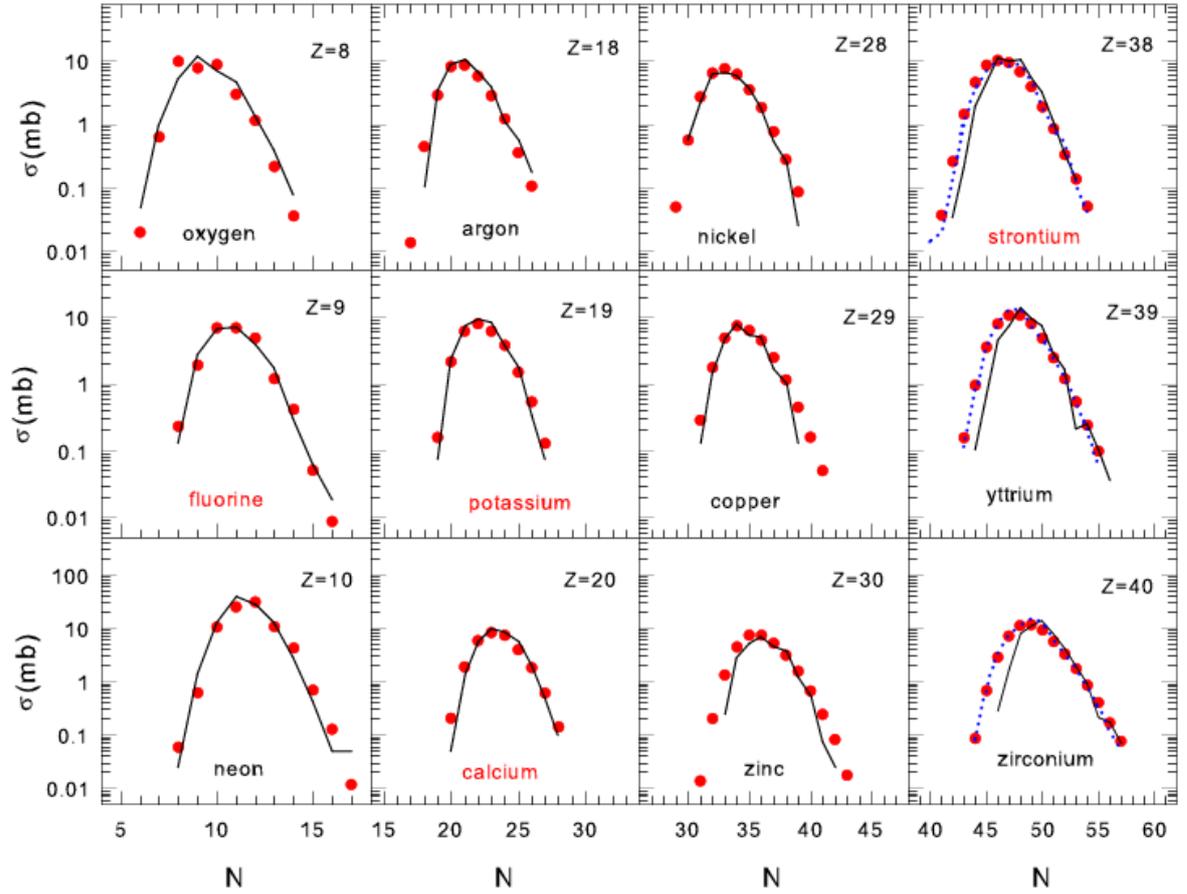

**Figure 2**. Production cross-sections of some selected isotopes measured in the reactions $^{136}$Xe + Pb at 1 GeV/nucleon of relativistic projectile energy. The red solid circles show the experimental data, and the solid lines the theoretically predicted results produced with ALADIN ensemble parameters. The blue dotted lines in the last coulumn show the compound contribution produced with the weakly excited compound nuclei ensemble. Isotope name in red refers to see the selected examples of some radionuclides including $^{18}$F, $^{40}$K, $^{41}$Ca, and $^{90}$Sr (see the text).

**Table 1:** The ensemble weighting parameters: Ex is the excitation energy per nucleon, Zs and As are the charge and mass number of the sources, respectively, N/Z is the neutron-to-proton ratio of the sources, and the last column is representing the statistical weight of the sources to optimize a good fit to the experimental data.

| $E_x$ (MeV/n) | Zs | As | N / Z | Weight |
| --- | --- | --- | --- | --- |
| 1 | 54 | 133 | 1.46 | 0.55 |
| 2 | 53 | 129 | 1.44 | 0.25 |
| 3 | 52 | 126 | 1.42 | 0.13 |
| 4 | 51 | 122 | 1.40 | 0.07 |